# TRB PAPER # 18-05532

# Impact of Vehicle-to-Vehicle Communication Reliability on Safety Applications: An Experimental Study

Mohammad A Hoque, PhD
Director, Vehicular Network Laboratory, East Tennessee State University &
Researcher, Center for Transportation Research, The University of Tennessee
hoquem@etsu.edu

Md Salman Ahmed
Researcher, Center for Transportation Research, The University of Tennessee &
Vehicular Network Laboratory, East Tennessee State University
ahmedm@etsu.edu

Jackeline Rios-Torres, PhD
Eugene P. Wigner Fellow, Oak Ridge National Laboratory
riostorresj@ornl.gov

Asad J. Khattak, Ph.D.
Beaman Professor, Department of Civil & Environmental Engineering
The University of Tennessee
akhattak@utk.edu

Ramin Arvin
Graduate Research Assistant, Department of Civil & Environmental Engineering
The University of Tennessee
rarvin@vols.utk.edu

**ACKNOWLEDGEMENTS**
This paper is based upon work supported by the US National Science Foundation under grant No. 1538139, and in part by the Laboratory Directed Research and Development Program of the Oak Ridge National Laboratory, Oak Ridge, TN 37831 USA, managed by UT-Battelle, LLC, for the DOE. Any opinions, findings, and conclusions or recommendations expressed in this paper are those of the authors and do not necessarily reflect the views of the sponsors.

Word Count: 1314

---

Submitted for Presentation Only to the 97th Annual Meeting
Transportation Research Board
January 2018
Washington, D.C.

## INTRODUCTION

Vehicle-to-vehicle (V2V) communication, such as Dedicated Short-Range communication (DSRC), can provide critical information and warnings with the aim to increase safety and reduce crashes. Simultaneous communication between vehicles can potentially reduce the traffic accidents and fatalities. In this study, we analyze the empirical data collected by equipping two vehicles with DSRC and driving them in opposite directions in city and highway environment *(1, 2)*. The city experiments were conducted near the business corridor of Johnson City, Tennessee, and the rural experiments were conducted on the I-26 interstate in Erwin town, Tennessee. Two On-Board Units (OBUs) provide single hop measurements to study the reliability of the V2V communication.

The "Do not Pass Warning" application has been listed as one of the prioritized safety applications by the United States Department of Transportation (USDOT) (*3*). The goal of this application is to detect an oncoming vehicle and provide warnings to the driver of the host vehicle to prevent an unsafe pass action. The success of such application is heavily dependent on the communication between the vehicles and any failure could endanger the driver safety. The objective of this research is to evaluate some of the possible V2V communication limitations and or failures that can be critical for the reliable operation of a "safe-pass advisory" application, through experimental testing under realistic conditions. In particular, we attempt to answer the following questions:

1- What are the impacts of the communication range and the reliability of DSRC on the effectiveness of specific Connected and Automated Vehicle (CAV) applications?
2- Which factors can affect the range of the communication and reliability of DSRC?

## METHODOLOGY

### Mathematical model for safe pass advisory application

In this section we develop a mathematical model to formulate the constraints on DSRC communication range and time. Figure 1 illustrates the scenario under investigation, i.e., a two-lane rural road at two instants of time: 1) the time that the eastbound car tries to pass the truck ($t_0$); and 2) the time that the passing is completed ($t_1$). We assume that the road grade is zero.

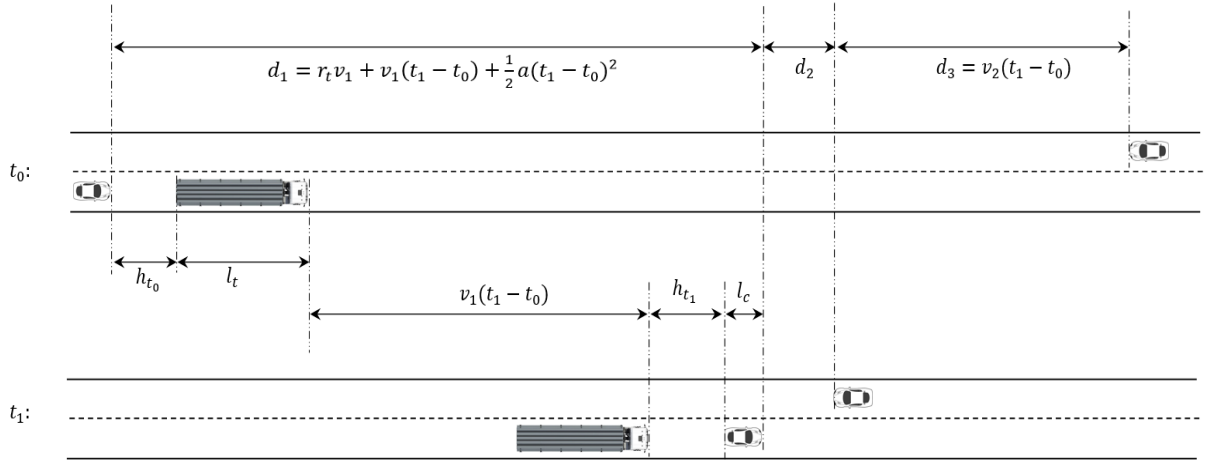

**FIGURE 1 Conceptual Diagram to derive the mathematical model of vehicle passing**

Where:
$v_1$: eastbound car and truck speed at $t_0$
$v_2$: westbound car speed at $t_0$
$r_t$: reaction time
$h_{t_0}$: space gap at $t_0$
$h_{t_1}$: space gap at $t_1$
$l_c$ = vehicle length
$l_t$ = truck length
$d_1$: distance covered by the eastbound car to complete passing the truck
$d_2$: minimum safety distance required to avoid collision
$d_3$: distance covered by the westbound car to allow passing

We can write:
$$d_1 + d_2 + d_3 < x_{v_1, v_2} \tag{1}$$
where, $x_{v_1,v_2}$ is the maximum range of DSRC communication for 2 vehicles approaching each other at $v_1$ and $v_2$ speeds . To avoid head-on collision:
$$t_v \geq \sqrt{2(2h_v + l_c + l_t - r_t v)\big/a_{max}^v} \tag{2}$$

$$x_v \geq (2h_v + l_c + l_t) + d_2 + 2v\sqrt{2(2h_v + l_c + l_t - r_t v)\big/a_{max}^v} \tag{3}$$

Thus, the successful implementation of the "safe pass advisory" application is subject to a time constraint (equation (2)) and a communication range constraint (equation (3)) for DSRC.

To determine the lower band for equations (2) and (3), a two-lane highway with speed limit of 55 mph was considered. The following assumptions are made:
$l_c = 5$ m
$l_t = 20$ m (length of a truck)

$a_{max}^{55} = 0.67$ m/s$^2$ (maximum acceleration at 55 mph) *(4, 5)*
$h_v = 24.6$ m (considering 1 second headway between the vehicle and the truck)
$r_t = 1\ s$ (driver reaction time)
$d_2 = 40$ m
Using these values, the lower band for time and range of the communication is: $t_v \geq 12.16$ s and $x_v \geq 712.4\ m$.

**Field Experimentation**

To determine the V2V communication range and time, we carried out V2V experiments using two vehicles in a real highway environment. Prior to conducting the experiment, official approval from Institutional Review Board (IRB) was obtained. The vehicles were equipped with after-market DSRC on-board units which are manufactured by ARADA Systems.

**RESULTS FROM FIELD TESTS**

**Impact of OBU Placement**

1) OBU placed inside the vehicles. The experimental data showed that there is a significant difference in the communication range and time when the OBU is mounted on rooftop vs. inside the car. Although the OBU has the omnidirectional antenna, the communication range was more than double for forward direction in comparison with backward direction when the OBU was placed inside the car. When the vehicles are approaching each other at 55 mph (or 70 mph), the communication range is 466 meters (or 401 meters). When the OBU is placed inside the vehicle, the backward communication depends on the OBU placement and vehicle type and the data propagate through various interior obstacles. Thus, the results are far worse in comparison with forward communications (327 meters for 55 mph and 400 for 70 mph).

2) OBU mounted on rooftop: in this case, the communication range for both forward and backward communication is almost the same and is about 1 km. There is just a slight increment in range for reverse direction which is due to the orientation of the antenna.

The results from both placements show that even with multi-hop V2V communication it is not feasible to design a "Safe Pass Advisory" application unless an outdoor DSRC antenna is used to maximize the communication range in forward and backward direction. This result only applies to the specific technology tested in the experiment. The subsequent experiments focused on the rooftop OBU.

**Duration of Connectivity between Opposite Vehicular Traffic**

The total communication period, or the duration of V2V connectivity between opposite vehicular traffic determines the amount of information that can be relayed to a vehicle in the opposite direction. An example use case of this would be a "detour advisory" information that is relayed to the upstream traffic heading towards a congestion due to accident on freeway. Figure 2 describes the duration of connectivity between opposite traffic in a highway environment. The experimental results suggested that it is possible to desing and implement a reliable "Safe pass advisory" application if the DSRC transceiver is mounted at the roof of the vehicle.

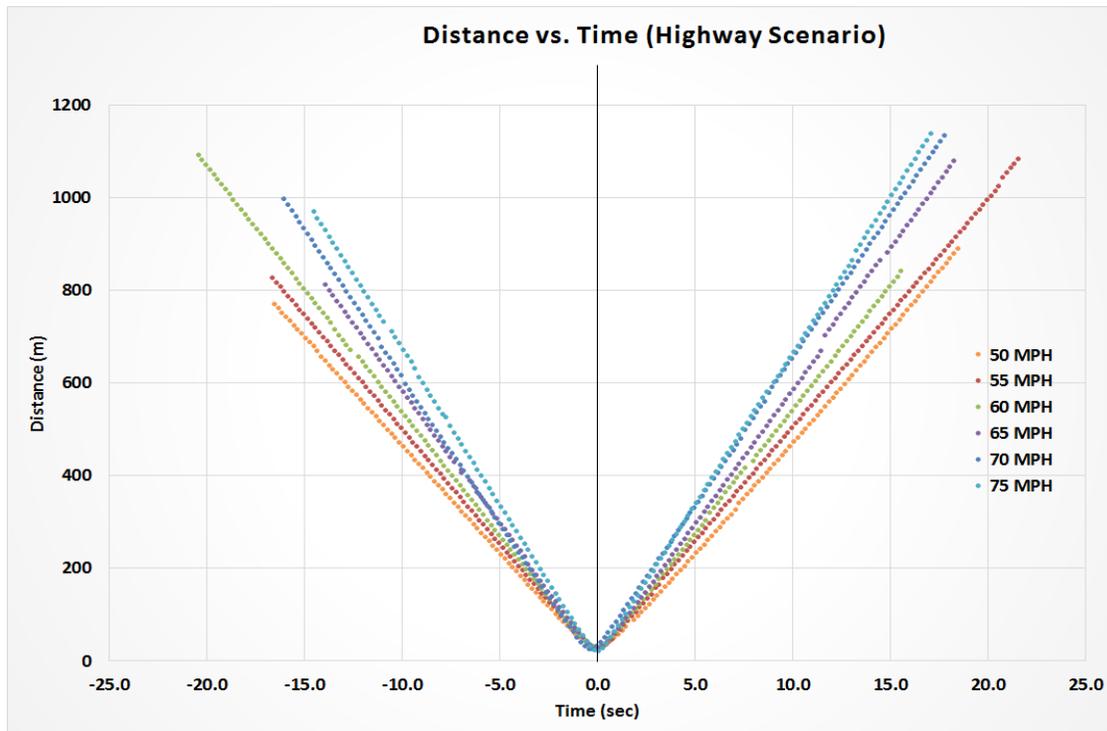

**FIGURE 2 Duration of Connectivity and range for highway (with rooftop OBU)**

**CONCLUSIONS**

The results from freeway experiments indicate that the communication range and connection period between two vehicles driven from opposite direction is more than doubled if the OBU is mounted on rooftop instead of placing it inside the vehicle (which is the usual case). Alternatively, if the OBU is placed inside the vehicle the communication range towards the backward direction is less than one-half of the range achieved in the forward direction. The reason behind this phenomenon is that, when the signal propagates in forward direction, it only goes through the windshield facing less obstacles. On the contrary, when the signal propagates towards the back direction, it has to find its way through the backseats and trunk/cargo spaces struggling with more obstacles. However, if the OBU (or just the DSRC antenna) is mounted on the roof, then the effective communication range is almost the same along every direction surrounding the vehicle.

The experimental results from city environment provides us some crucial information about how the V2V communication is obstructed by the change of altitude within a straight road segment, even with a rooftop OBU. This translates into additional constraints for designing a safe pass advisory application for two-lane rural highways that have frequent change of altitudes. In summary, the contribution of this paper are as follows:

1. Developing a mathematical analysis to understand the required V2V communication parameters and constraints pertaining to a DSRC-based "Safe pass advisory" application for two-lane rural highways.
2. Experimental quantification of V2V communication range and connectivity period between two DSRC-equipped vehicles approaching each other from opposite directions at various speeds, for both freeway and city environment. The data obtained from this experiment helped to determine the time and distance constraints for the above mentioned "Safe pass advisory" application.
3. Investigation of the impact of varying altitude on V2V communication reliability and its implications on safety-critical applications.
4. Evaluation of the impact of vehicle-interior obstacles and OBU placement on the reliability, range and duration of V2V communication, both in forward and back direction. The collected experimental data provided insights for utilizing multi-hop V2V communication to overcome the limitations of the "Safe pass advisory" application.